\documentclass[10pt,journal,compsoc]{IEEEtran}

\usepackage[export]{adjustbox}
\usepackage{wrapfig}
\usepackage{graphicx}
\graphicspath{ {./figs/} }
\usepackage{comment}
\usepackage{array}
\usepackage{amsmath, amssymb, amsthm, bm}
\usepackage{setspace}
\usepackage{multicol}
\usepackage{lipsum}
\usepackage{changepage}
\usepackage{textcomp}
\usepackage{enumitem}
\usepackage{color}
\usepackage{xurl}
\usepackage{tikz}
\usetikzlibrary{positioning}
\usetikzlibrary{arrows.meta}
\usepackage[font=footnotesize, labelfont=bf, textfont=bf]{caption}
\usepackage{subfig}
\usepackage{epsfig}
\usepackage{svg}
\usepackage{rotating}
\usepackage{multirow}
\usepackage{makecell}
\usepackage{booktabs}
\usepackage{longtable}
\usepackage{threeparttable}
\usepackage{tabularx}

\usepackage{lscape}
\usepackage[ruled,vlined,linesnumbered]{algorithm2e}
\SetKwInOut{Input}{input}
\SetKwInOut{Output}{output}
\usepackage{listings}
\usepackage{verbatim}
\usepackage[round]{natbib}
\usepackage[american]{babel}
\usepackage[autostyle, english=american]{csquotes}
\MakeOuterQuote{"}
\usepackage[colorlinks=true]{hyperref}
\hypersetup{
    colorlinks=true,
    linkcolor=blue,
    citecolor=blue,
    filecolor=magenta,
    urlcolor=blue,
    pdftitle={Overleaf Example}
}
\usepackage{cleveref}

\crefalias{proposition}{proposition}

\usetikzlibrary{shapes.geometric, arrows.meta, positioning}

\begin{document}

\captionsetup[table]{labelformat={default},labelsep=period, name={Table}}

\captionsetup[figure]{labelformat={default},labelsep=period, name={Figure}} 

\title{Spatial Two-Stage Hierarchical Optimization Analysis for Site Selection of Bitcoin Mining in South Korea}

\author{
    \IEEEauthorblockN{
        Yoonseul Choi\textsuperscript{1}, 
        Jungsoon Choi\textsuperscript{2, 3}
    }\\
    \IEEEauthorblockA{
        \textsuperscript{1}Department of Applied Statistics, Hanyang University, Seoul, South Korea
    }\\
    \IEEEauthorblockA{
        \textsuperscript{2}Department of Mathematics, Hanyang University, Seoul, South Korea
    }\\
    \IEEEauthorblockA{
        \textsuperscript{3}Research Institute for Natural Sciences, Hanyang University, Seoul, South Korea
    }\\    
    \IEEEauthorblockA{
        Emails: \href{mailto:clodagh@hanyang.ac.kr}{clodagh@hanyang.ac.kr},
        \href{mailto:jungsoonchoi@hanyang.ac.kr}{jungsoonchoi@hanyang.ac.kr}
    }
}

\markboth{Y. Choi \& J. Choi: Spatial Two-Stage Hierarchical Optimization Analysis for Site Selection of Bitcoin Mining in South Korea}%
{Y. Choi \& J. Choi: Spatial Two-Stage Hierarchical Optimization Analysis for Site Selection of Bitcoin Mining in South Korea}

\IEEEtitleabstractindextext{%
\begin{abstract}
South Korea faces the dual challenge of managing growing distributed solar energy surpluses and the high energy demand of industries like Bitcoin mining. Leveraging mining operations as a flexible load to monetize this `net-metering surplus' presents a viable synergy, but requires a robust site selection methodology. Traditional GIS-based Multi-Criteria Decision Analysis (MCDA) struggles with subjective weighting and integrating heterogeneous spatial data (areal-level and lattice-level). This thesis develops and implements a Two-Stage Hierarchical Optimization framework to overcome these limitations. Stage 1 (Areal-Level) employs a cost-benefit optimization to determine the optimal number ($K^*$) and combination of regions, maximizing a final adjusted net profit by balancing surplus power revenue against detailed land and non-linear infrastructure costs. Stage 2 (Point-Level) then uses a GIS-based sliding window search within these selected regions, applying topographic (slope $< 6.0^\circ$) and land-use constraints at a 30m resolution to identify physically constructible `unit sites'. The model identified an optimal configuration of $K^*=3$ regions (Yongin, Damyang, Miryang) yielding a maximum potential net profit of approximately \$307 million. Crucially, the Stage 2 screening revealed that Yongin, the most profitable region, was also the most physically constrained, 87\% of sites filtered out. This research contributes a scalable, objective framework for energy infrastructure siting that effectively integrates multi-scale spatial data. It provides a data-driven strategy for policymakers and grid operators (like Korea Electric Power Corporation) to monetize curtailed renewables and enhance grid stability.
\end{abstract}
\vfill

}

\maketitle
\IEEEdisplaynontitleabstractindextext
\IEEEpeerreviewmaketitle

\section{Introduction}
\label{sec:ch1}

\subsection{Background}

The global energy system is undergoing a structural transformation driven by decarbonization, decentralization, and digitalization \citep{CAF3DEnergy, GreenvoltDecentralized, WIPODER2025}. As these forces reshape how electricity is produced, transmitted, and consumed, the global energy transition toward renewable and sustainable systems presents both formidable challenges and promising opportunities \citep{PSURenewableChallenges, PlainConceptsFutureEnergy}. In particular, the widespread deployment of distributed solar generation has introduced temporal and regional mismatches between supply and demand, which frequently result in large volumes of curtailed or unused surplus electricity \citep{RatedPowerCurtailment}. For example, California has experienced significant solar curtailment in recent years, with a 29\% increase in curtailed utility-scale wind and solar output in 2024, totaling 3.4 million MWh \citep{eia2023}. Such curtailment highlights the need for smarter, more flexible grid solutions to reduce energy loss and avoid unnecessary transmission upgrades \citep{eia2023}. This phenomenon is especially prominent in countries with rapidly expanding renewable energy portfolios. In South Korea, renewable energy installed capacity and power generation have been steadily increasing, with particularly notable growth centered on solar power generation \citep{KNEA_Supply_Stats}. However, grid bottlenecks are emerging as renewable energy capacity rapidly expands \citep{IEEFAKoreaBottlenecks, ChosunSolarShutdown}. Similarly, in Germany, solar curtailment jumped by 97\% in 2024 \citep{pvmag_germany_2025}. China has also seen rising renewable energy curtailment as rapid solar and wind expansion outpaces grid capacity. In the first half of 2025, solar and wind curtailment rose to 6.6\% and 5.7\% \citep{reuters_china_curtaiment_2025}.

In parallel, the digital economy has seen the emergence of highly energy-intensive industries, among which the mining of Bitcoin stands out \citep{kufeoglu2019energy, ccaf2025mining}. Through its Proof-of-Work\footnote{The Proof-of-Work (PoW) consensus mechanism, used by cryptocurrencies like Bitcoin, requires network participants (miners) to use substantial computational power to solve complex mathematical puzzles in order to validate transactions and add new blocks to the blockchain \citep{Nakamoto2009Bitcoin}.} consensus mechanism, the Bitcoin network consumes electricity at the scale of a mid-sized country, raising serious questions about sustainability and grid impact \citep{alex2020, rmi2023, JBSSustainableBitcoin2025}. The combination of high energy consumption with a globally distributed computational infrastructure places mining at a critical intersection of energy, policy, and economics \citep{lal2024jcp, HutabaratBitcoinRenewable}.

However, this high intensity of consumption contains within it a potential synergy with the renewable energy challenge \citep{khosravi2023}. Strategically located bitcoin mining operations can act as flexible consumers that absorb surplus renewable power, converting otherwise wasted electricity into digital economic value, while providing grid services by stabilizing demand where and when generation is abundant \citep{HutabaratBitcoinRenewable}. Several research suggest that mining can help mitigate curtailment and contribute to more efficient utilization of power infrastructure \citep{carter2023, DASAKLIS2025100114}.

This dual nature, heavy electricity consumption on the one hand, and the potential to serve as a flexible load tied to renewable surplus on the other, can integrate with a spatial optimization problem with important policy and economic implications \citep{NIAZ2022132700}. In South Korea, distributed-solar installations under net-metering arrangements generate surplus electricity when on-site consumption is exceeded by generation, resulting in so-called ``net-metering surplus power'' that can be supplied back to the grid and offset against consumption \citep{IEA2020_KoreaNetMetering}. Under the net-metering (or ``net energy metering'') scheme, a solar-owner uses the generated electricity onsite first, and any remaining surplus quantity is fed into the grid and credited against future electricity use \citep{KEAJ3714, DSIREGlossary}.

\begin{figure}[h]
\centering
\includegraphics[width=0.45\textwidth]{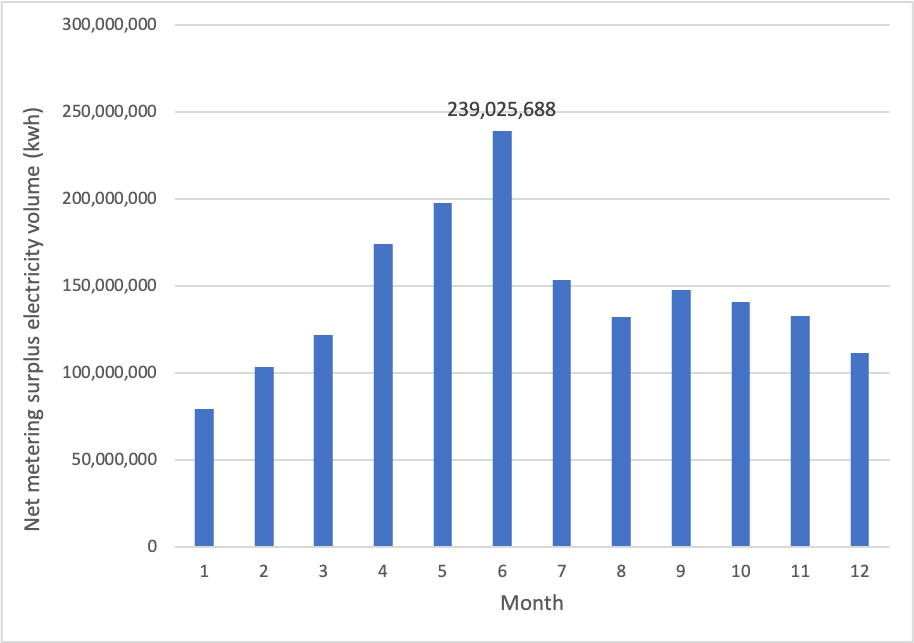}
\caption{Surplus electricity in South Korea, 2024}
\label{fig:surplus2024korea}
\end{figure}

In the context of Korea Electric Power Corporation\footnote{55, Jeollyeok-ro, Naju-si, Jeollanam-do, South Korea, available at \url{https://home.kepco.co.kr/kepco/EN/main.do}} (KEPCO)’s service area, this surplus can represent both a resource and a challenge. For example, Figure \ref{fig:surplus2024korea} shows the net-metering surplus electricity volumn (kWh) in South Korea in 2024.
The surplus electricity of 239,025,688 kWh in June of 2024 is equivalent to the annual consumption of around 550,000 households in South Korea, assuming a typical household uses 360 kWh per month \citep{donga2024}. Moreover, the surplus electricity often accumulates in certain periods or regions and may be under-utilized if conventional means of consumption or storage are lacking \citep{conversation2025renewable}. For this reason, leveraging this surplus through high-demand flexible loads, such as Bitcoin mining operations, appears particularly attractive \citep{rudd_porter_2024}.

The substantial electricity consumption of Bitcoin mining has highlighted the potential for utilizing surplus or curtailed electricity from renewable energy generation. A recent study by \citet{choi2025} demonstrated that using KEPCO’s surplus electricity for Bitcoin mining is feasible and profitable, and could help KEPCO improve its energy‐resource efficiency, reduce debt burdens and monetise otherwise wasted solar production. By focusing on KEPCO’s solar net-metering surplus as the primary energy supply base for potential mining site selection, this research draws direct policy linkage between renewable surplus management and demand-side optimisation. \citet{dorrell2023green} analyzed the profitability of using excess wind energy. According to the results of this study, when utilizing wind power plant curtailment, Bitcoin mining could provide maximum investment returns of 1,347\% and 805\% when starting on November 1, 2021, and November 1, 2022, respectively. \citet{shan2019caiso} simulated deploying miners at renewable plants on the California Independent System Operator (CAISO) and reported that the CAISO system will earn \$5.6--48.1 million annual net profit with 50.8--79.9\% curtailment absorption in 2018. \citet{lal2024jcp} demonstrated the economic advantage of Bitcoin mining compared to $H_2$ production using surplus power. Their results indicate that mining is profitable in 80 out of 83 examined projects, achieving up to \$7.68 million in profit at a site and utilizing up to 62\% of available renewable output.

In the case of South Korea, identifying suitable sites for bitcoin mining requires the integration of multiple spatial and economic criteria: regional surplus electricity availability, land cost and accessibility, topographical feasibility (e.g., slope), and connectivity to grid infrastructure. Each of these variables resides at different spatial resolutions. Surplus electricity data tend to be measured at regional or administrative levels, whereas physical terrain and land parcel characteristics are recorded at much finer, grid-level scales.

\subsection{Preliminaries}
Conventional Geographic Information Systems (GIS) combined with Multi-Criteria Decision Analysis (MCDA) frameworks have been widely used in the siting of renewable power plants and related infrastructure \citep{su142214742, su15108359, ISLAM2024119595}. MCDA is a structured decision-making framework that evaluates alternatives based on multiple, often conflicting criteria, enabling transparent and systematic prioritization in complex planning problems \citep{saaty1980, greco2016mcda}. This methodology also involves overlaying various spatial layers (e.g., slope, land use, weather) and assigning weights to each layer to derive a final `suitability index' map. Within MCDA, the Analytic Hierarchy Process (AHP) is a structured decision-making framework for complex problems that involves breaking a decision into a hierarchy of goals, criteria, and alternatives \cite{forman2001analytic}. Also, its integration with Geographic Information Systems (GIS–AHP) allows spatially explicit decision-making by combining expert judgments with geospatial datasets for site selection and regional planning. Technique for Order Preference by Similarity to Ideal Solution (TOPSIS) is used to compare and rank a set of alternatives based on multiple criteria. The core idea is that the chosen alternative should have the shortest distance from the positive ideal solution (the theoretical best) and the longest distance from the negative ideal solution (the theoretical worst) \citep{hwang1981multiple}. Extensions such as Fuzzy AHP incorporate fuzzy set theory to better handle ambiguity and uncertainty inherent in expert assessments, improving the robustness of weighting processes \citep{chang1996fuzzyahp}. Similarly, Fuzzy TOPSIS adapts the classical Technique for Order Preference by Similarity to Ideal Solution by representing criteria evaluations with fuzzy numbers, enabling more realistic ranking of alternatives when decision environments involve linguistic or uncertain information \citep{chen2000fuzzytopsis}. These MCDA approaches provide flexible and rigorous tools for evaluating renewable energy siting, grid planning, and infrastructure optimization under uncertainty.

\citet{ISLAM2024119595} utilized GIS-AHP for solar power plant siting in Bangladesh, weighting criteria such as solar irradiation, temperature, slope, land use, and proximity to urban areas, roads, power lines to generate a suitability map. Similarly, \citet{su142214742} used GIS-AHP for wind farm site selection in Khuzestan, Iran, evaluating 14 criteria including wind speed, slope, and proximity to airports, power lines, roads, residential areas to identify optimal locations. \citet{gupta2022} also applied GIS-MCDA-AHP for flood risk assessment in Assam, India, developing a risk index by integrating various indicators like elevation, slope, rainfall, and population density. \citet{Somasi_Kondamudi_2024} employed Fuzzy AHP and Fuzzy TOPSIS for selecting sites for solar desalination in Visakhapatnam, India, evaluating technical, economic, environmental, social, and political factors. Meanwhile, \citet{su15108359} proposed a approach to address the location-dependent evaluation problem of traditional MCDA. They used GIS-AHP to define global classification bounds for criteria and generated a synthetic dataset through permutations of these bounds. Subsequently, they constructed a multiple linear regression model where the `site score' (calculated using AHP weights) served as the dependent variable, and the criteria values (solar irradiation, slope, distances, etc.) acted as independent variables. This model aims to predict a global suitability score for any given location based solely on its criteria values, attempting to overcome the subjectivity and local dependency of MCDA.

While GIS-based MCDA methods are useful for integrating heterogeneous spatial factors into a single suitability surface, they have two important limitations for the present study. First, methods such as AHP necessarily rely on expert judgement to assign weights, so the resulting suitability index is not an objective guarantee of an optimal allocation under explicit resource or budget constraints. Second, a suitability map alone does not produce an implementable solution when the decision problem requires discrete facility counts, capacity limits, or system-level constraints (e.g., total available surplus electricity across regions). 

To address these limitations, previous work has combined GIS data with mathematical optimization or statistical models. In particular, linear programming (LP) can identify cost- or capacity-optimal facility deployments by optimizing an explicit objective subject to resource constraints \citep{church2009location, murray2010gis, malczewski2015gis}. LP is a mathematical optimization technique that assumes a linear relationship among decision variables, the objective function, and the constraints. In spatial applications, LP can incorporate costs, capacities, distances, and demand requirements to select facility locations efficiently \citep{dantzig1963linear}. Compared with MCDA methods alone, LP provides an explicit quantitative framework to identify optimal solutions under well-defined constraints \citep{malczewski2015gis}. 

However, applying existing methods to this study presents difficulties. Since there is no prior research identifying the extent to which each explanatory variable influences the dependent variable, it is not possible to obtain or estimate empirical data for the dependent variable itself. In particular, in studies concerning the optimal location of Bitcoin mining operations, this limitation can be more evident. Since there is little empirical research identifying the key variables that influence Bitcoin mining output or profitability, the analysis falls into this category, making it difficult to determine suitable locations with confidence. Even if mining revenue is used as the dependent variable, research on how mining revenue is directly affected by factors like temperature or humidity is lacking, making it difficult to handle via linear programming.

In response to these difficulties, this study proposes a spatial two-stage hierarchical approach for the optimal site selection of bitcoin mining farms across South Korea. The proposed framework proceeds in two stages. First, an LP formulation determines the optimal number and approximate locations of deployment centers under constraints of surplus electricity availability and cost thresholds. LP is used here because it explicitly encodes objective functions and resource constraints and thus yields system-level optimal allocations. Second, in the first stage, the number of regions and approximate regions were determined, so within each region, lattice-level screening (e.g., slope) to produce implementable site lists. Therefore, GIS-based analysis refines the feasible zones by integrating areal-level (regional surplus energy, official land price) and fine lattice-level (topographic slope) spatial layers. This hierarchical separation, from areal economic optimization to lattice-level physical screening, (i) aligns each analytical step with the spatial scale at which its inputs are best supported, (ii) reduces reliance on subjective weightings by using LP where quantitative constraints are available, and (iii) substantially reduces computational burden by restricting fine-resolution search to LP-identified feasible areas.

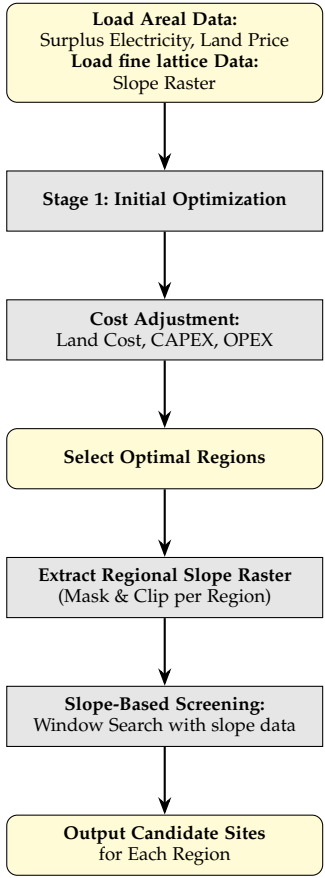
\begin{figure}[h]
  \centering
  \begin{tikzpicture}[
    node distance=0.9cm,
    every node/.style={font=\scriptsize, align=center},
    startstop/.style={rectangle, rounded corners, draw, fill=yellow!20,
                      minimum width=4.2cm, minimum height=0.8cm},
    process/.style={rectangle, draw, fill=gray!20,
                    minimum width=4.2cm, minimum height=0.8cm},
    arrow/.style={-{Stealth}, thick}
  ]
    \node (input) [startstop]
      {\textbf{Load Areal Data:}\\ Surplus Electricity, Land Price \\
      \textbf{Load fine lattice Data:}\\ Slope Raster};
    \node (stage1a) [process, below=of input]
      {\textbf{Stage 1: Initial Optimization}};

    \node (stage1b) [process, below=of stage1a]
      {\textbf{Cost Adjustment:}\\ Land Cost, CAPEX, OPEX};

    \node (stage1c) [startstop, below=of stage1b]
      {\textbf{Select Optimal Regions}};

    \node (stage2a) [process, below=of stage1c]
      {\textbf{Extract Regional Slope Raster}\\ (Mask \& Clip per Region)};

    \node (stage2b) [process, below=of stage2a]
      {\textbf{Slope-Based Screening:}\\ Window Search with slope data};

    \node (final) [startstop, below=of stage2b]
      {\textbf{Output Candidate Sites}\\ for Each Region};

    \draw [arrow] (input)  -- (stage1a);
    \draw [arrow] (stage1a) -- (stage1b);
    \draw [arrow] (stage1b) -- (stage1c);
    \draw [arrow] (stage1c) -- (stage2a);
    \draw [arrow] (stage2a) -- (stage2b);
    \draw [arrow] (stage2b) -- (final);

  \end{tikzpicture}
  \caption{Two-Stage Hierarchical Optimization Workflow}
  \label{fig:analysis_flow_two_stage}
\end{figure}

In sum, as seen in the Figure \ref{fig:analysis_flow_two_stage}, the two-stage hierarchical approach combines the normative strengths of LP with spatially explicit GIS screening to produce both objective, constraint-aware regional allocations and realistic, implementable point-level site candidates. This strategy mitigates potential modifiable areal unit problem\footnote{Modifiable Areal Unit Problem (MAUP) refers to the issue where statistical analysis results (e.g., correlation coefficients, regression coefficients) vary significantly depending on the scale of spatial units used for aggregation and the zoning method, even when using the same underlying spatial data \citep{openshaw1983modifiable}.} and parameter uncertainty while retaining computational tractability and interpretability for Bitcoin mining site selection. Finally, this thesis contributes a spatial multilevel LP workflow that: (i) uses areal surplus electricity and land price as LP inputs to derive optimal regional capacities and the number of deployment centers, and (ii) refines those regional candidates using slope at point resolution to produce implementable site lists. By doing so, it translates optimization strengths into the GIS–MCDA context.

The remainder of this thesis is organized as follows.  Section~\ref{sec:ch2} details the proposed methodology, areal-level modeling and slope-based lattice-level refinement. Section~\ref{sec:ch3} presents data sources, parameter settings, and experimental results, followed by discussion and policy implications. Finally, Section~\ref{sec:ch4} concludes the study with key findings and future research directions.

\section{Methodologies}
\label{sec:ch2}

To derive the optimal sites for Bitcoin mining, this study proposes a Two-Stage Hierarchical Optimization framework, which combines macro-level site selection at the areal unit (sigungu\footnote{Sigungu is a collective term for the basic local administrative units in South Korea, comprising Si (city), Gun (county), and Gu (autonomous district). In this study, the Sigungu serves as the primary areal unit for the aggregation and analysis of macro-level variables (e.g., surplus electricity, land price) in the Stage 1 optimization.}) ith micro-level site placement at the finer lattice (grid) unit.

\subsection{Overview}
This study adopts Two-Stage Hierarchical Optimization framework to integrate heterogeneous data of different spatial scales (areal and lattice) and ensure computational feasibility. As optimizing all pixels across the nation simultaneously is computationally infeasible, the analysis is divided into two parts; macro and micro levels. \textbf{Stage 1} utilizes areal-level data (e.g., surplus electricity, official land price) to select $K$ optimal regions (e.g., sigungu units) with the highest potential. \textbf{Stage 2} then identifies the specific optimal sites (pixels) for facility placement within those selected regions, considering point-level constraints such as terrain slope.

\subsection{First Stage: Areal-Level Optimal Region Selection} \label{sec:stage1_lp}
The objective of Stage 1 is to identify the optimal combination of $K$ sigungu regions ($j$) across the nation that yields the highest economic feasibility. This is structured as a two-sub-step process.

\subsubsection{Initial Optimization}
First, an initial optimization is performed based on surplus electricity data and Bitcoin network data. This step determines the installable number of miners ($N_j$) for each region $j$, expected annual revenue, and annual depreciation for various scenarios of $K$ combined regions ($K=1, ... , 5$). These initial optimization results are structured into a detail by dictionary object, which serves as the input for the secondary analysis.

\subsubsection{Full Cost-Benefit Analysis}
The profit result from the initial optimization does not fully account for land acquisition and infrastructure development costs. Therefore, an algorithm is required to incorporate these real-world costs and calculate a fully-adjusted net profit. This process involves the
estimation of the following key cost components.

\paragraph*{Annualized Land Cost:} The total required size of land area ($A_j$) for each region ($j$) is first determined based on a container-based approach, where modular mining units are housed in standardized shipping containers. The total land cost $L_j$ of each region $j$ is computed as:
$$L_j = \frac{A_j * P_j * M_{\mathrm{market}}}{L_{\mathrm{amort}}}$$
where $A_j$ is the required area, $P_j$ is the average official land price (KRW/$\mathrm{m^2}$), $L_{\mathrm{amort}}$ is the land amortazation and $M_{market}$ is the market multiplier\footnote{The market multiplier reflects the discrepancy between official and market prices. In this study, $M_{\mathrm{market}}$ = 1.3 \citep{FRANGIAMORE2025107012}.}. The total annualized land cost ($L_K$) for a $K$ regions is:
$$
L_K = \sum_{j \in \mathbf{Y}_K} L_j
$$
while $\mathbf{Y}_K$ is the set of candidate regions.

\paragraph*{Annualized Infrastructure Cost:} The infrastructure cost ($I_K$) consists of capital expenditures (CAPEX) and fixed operational expenditures (OPEX):
$$
I_K = \frac{C_{\mathrm{capex}}(\mathrm{MW}_K, K)}{Y_{\mathrm{infra}}} + K \times O_{\mathrm{fixed}}
$$
where $C_{\mathrm{capex}}(\mathrm{MW}_K, K)$ represents the total installation cost based on total power capacity ($\mathrm{MW}_K$) and number of sites ($K$), annualized over the infrastructure lifetime ($Y_{\mathrm{infra}}$) and fixed OPEX costs required per region ($O_{\mathrm{fixed}}$).

\paragraph*{Annual Energy Cost:} As this model is predicated on the utilization of surplus electricity, the base energy price is set to $0$, assuming zero electricity cost \citep{dorrell2023green}.

\paragraph*{Algorithm Overview:} 
Algorithm \ref{alg:stage1} summarizes the iterative process for selecting the optimal number of regions $K$ and corresponding set $\mathbf{Y}$ that maximize the adjusted annual profit $\prod_{\mathrm{adj}}$ with number of total miners ($N_k$) after accounting for all annualized costs. \\

\begin{algorithm}[ht]
\small
\caption{Stage 1: Areal-Level Optimal Region Selection}
\label{alg:stage1}
\SetAlgoLined
\KwIn{
\begin{itemize}
    \item Initial optimization results $\mathcal{D}_{\text{init}} = \{K \rightarrow (\mathbf{Y}_K, \mathbf{N}_K, \Pi_{\text{orig}})\}$ 
    \item Areal land price data $\mathcal{P}_{\text{land}}$ (KRW/m$^2$)
    \item Power per machine $P_{\text{machine}}$
    \item Land cost parameters $\mathcal{C}_{\text{land}}$ ($A_j$, $M_{\text{market}}$, $L_{\text{amort}}$)
    \item Infrastructure parameters $\mathcal{C}_{\text{infra}}$ ($C_{\text{capex}}$, $O_{\text{fixed}}$, $Y_{\text{infra}}$)
\end{itemize}
}
\KwOut{Optimal number of sites $K^*$, optimal region set $\mathbf{Y}^*$, maximum adjusted profit $\Pi_{\text{max}}$}
\vspace{2mm}

Initialize $\Pi_{\text{max}} \leftarrow -\infty$, $K^* \leftarrow 0$, $\mathbf{Y}^* \leftarrow \emptyset$\;

\For{each candidate number of regions $K$ in $\mathcal{D}_{\text{init}}$}{
    Retrieve initial result $(\mathbf{Y}_K, \mathbf{N}_K, \Pi_{\text{orig}})$\;

    \tcc{1. Compute Annualized Land Cost}
    $L_K \leftarrow 0$\;
    
    \For{each region $j \in \mathbf{Y}_K$}{
        $p_j \leftarrow \mathcal{P}_{\text{land}}[j]$\;
        
        $L_j \leftarrow A_j * p_j * M_{\text{market}} / L_{\text{amort}}$\;
        
        $L_K \leftarrow L_K + L_j$\;
    }

    \tcc{2. Compute Annualized Infrastructure Cost}
    $MW_K \leftarrow \mathbf{N}_K * P_{\text{machine}}$\;
    
    $I_K \leftarrow C_{\text{capex}}(MW_K, K)/Y_{\text{infra}} + K \times O_{\text{fixed}} $\;
    
    \tcc{3. Compute Adjusted Net Profit}
    $\Pi_{\text{adj}} \leftarrow \Pi_{\text{orig}} - (L_K + I_K)$\;

    \If{$\Pi_{\text{adj}} > \Pi_{\text{max}}$}{
        $\Pi_{\text{max}} \leftarrow \Pi_{\text{adj}}$\;
        
        $K^* \leftarrow K$\;
        
        $\mathbf{Y}^* \leftarrow \mathbf{Y}_K$\;
    }
}
\Return $K^*, \mathbf{Y}^*, \Pi_{\text{max}}$\;
\end{algorithm}

\subsection{Stage 2: Lattice-Level Candidate Site Generation}
\label{sec:stage2}
The optimal regions identified in Stage 1 ($\mathbf{Y}^*$) have only confirmed macro-level feasibility. The objective of the second stage is to derive the precise locations of constructible `candidate sites' within these regions that satisfy topographic constraints.

\paragraph*{Regional Slope Data Extraction:} First, for each selected region ($j$) from Stage 1, the national $30\text{m}$ resolution slope raster is masked and clipped using the region's polygon geometry via the \texttt{rasterio.mask} operation. This extracts the slope data ($slope_j$) relevant only to that region.

\paragraph*{Defining Physical Constraints and Search Window:} A maximum allowable slope ($\theta$) is defined as the physical limit for construction (e.g., $\theta = 6.0^\circ$). Furthermore, based on the required land size ($A_j$) for each region $j$ derived in Stage 1, a `unit site area' necessary for each size of facility is defined (e.g., $80\text{m} \times 80\text{m}$ is required for the $6,400\text{m}^2$ facility).

\paragraph*{Sliding Window Search:} A window of the defined unit size ($\mathrm{Width}_j \times \mathrm{Height}_j$) is used to search the regional slope raster ($slope_j$). The window moves at a set interval (e.g., $10\text{m}$), and at each position, the algorithm checks if all pixels within the window have a slope value below the maximum threshold ($\theta$).

\paragraph*{Deriving the Candidate Site Set ($\mathcal{F}_j$):} The top-left pixel coordinates ($i, k$) of any window that satisfies this slope constraint are identified as a `feasible candidate site'. These pixel coordinates are then converted back to real-world coordinates  using the raster's Affine transform. This process generates the final output of Stage 2: a `candidate site set' $\mathcal{F}_j = \{r_1, r_2, ..., r_m\}$ for region $j$, where each candidate is a polygon of size $\mathrm{Width}_j \times \mathrm{Height}_j$.

\paragraph*{Algorithm Overview:} 
Algorithm \ref{alg:stage2} summarizes the iterative process for selecting the optimal grid in each region $j$. In Stage 2, detailed site candidates that are actually constructable are generated within the macro-level optimal candidate areas ($\mathbf{Y}^*$) derived in Stage 1. Based on the required area size ($A_j$) for each region, a unit area per facility is defined. A sliding window of this size is moved at a fixed interval, checking whether the slope value of all pixels falls below the allowable slope ($\theta$). All window positions satisfying the condition are recorded as `constructable candidate sites'. The corresponding pixel coordinates are transformed into the actual coordinate system via Affine transformation, forming the final candidate site set $\mathcal{F}_j$. This algorithm enables the systematic acquisition of only topography-based, micro-level candidate sites within each region. \\

\begin{algorithm}[ht]
\small
\caption{Stage 2: Lattice-Level Candidate Site Selection}
\label{alg:stage2}
\SetAlgoLined
\KwIn{\\Optimal region set from Stage 1: $\mathbf{Y}^*$\\
Maximum allowable slope $\theta$\\
Facility unit site dimensions for each region: $(\mathrm{Width}_j,\mathrm{Height}_j)$\\
Sliding window stride $\Delta$}
\KwOut{Candidate site sets $\mathcal{F}_j$ for all regions $j\in\mathbf{Y}^*$}

\For{each region $j \in \mathbf{Y}^*$}{
  \tcc{1. Regional slope raster extraction (mask + clip)}
  $slope_j(u,v)\leftarrow S(u,v)\cdot \mathbf{1}_{(u,v)\in\text{Polygon}(j)}$\;

  \tcc{2. Retrieve required site footprint}
  $(\mathrm{Width}_j,\mathrm{Height}_j)$ determined by $A_j$\;

  \tcc{Initialize}
  $\mathcal{F}_j \leftarrow \varnothing$\;

  \tcc{3. Sliding window search}
  \For{$i=0,\Delta,2\Delta,\dots$}{
    \For{$k=0,\Delta,2\Delta,\dots$}{
      \tcc{Windowed slope values:
      $S_{i,k}=\{\,slope_j(u,v)\mid i\le u<i+\mathrm{Width}_j,\;k\le v<k+\mathrm{Height}_j\,\}$}
      \If{$\displaystyle \max_{(u,v)\in S_{i,k}} slope_j(u,v) < \theta$}{
        \tcc{4. Convert pixel indices to world coordinates}
        Let $(x_{ul},y_{ul})$ be world coordinate of pixel $(i,k)$ computed by
        $ \begin{pmatrix} x_{ul}\\ y_{ul} \end{pmatrix} =
          A \begin{pmatrix} i\\ k\\ 1 \end{pmatrix}$, where $A$ is the affine geotransform matrix.\;

        \tcc{5. Define polygon in real-world coordinates}
        Candidate site:
        $r=\{(x,y)\mid x_{ul}\le x < x_{ul}+W_j,\; y_{ul}\le y < y_{ul}+H_j\}$\;

        \tcc{Add site}
        $\mathcal{F}_j \leftarrow \mathcal{F}_j \cup \{r\}$\;
      }
    }
  }
}
\Return $\{\mathcal{F}_j : j\in\mathbf{Y}^*\}$\;
\end{algorithm}

\section{Data Analysis}
\label{sec:ch3}

In this chapter, we apply the two-stage hierarchical optimization framework described in Section 3 to the actual spatial datasets of South Korea. All analyses were conducted under Apple M2 chip with 16GB of RAM. For software, the system ran on Sonoma 14.1.1, and all experiments were implemented in \texttt{Python} (3.10.13) environment, utilizing libraries such as \texttt{matplotlib} (3.6.1), \texttt{geopandas} (1.0.1), and \texttt{rasterio} (1.4.3).

\subsection{Data Source}
This study integrates four datasets to construct the spatial multilevel framework for national-scale Bitcoin mining site selection: (1) regional surplus electricity, (2) official land price, (3) Bitcoin network data, and (4) topographic slope data. Each dataset represents a distinct spatial or economic dimension of the proposed model.

\subsubsection{Surplus Electricity Data}
\label{sec:surpluselectricity}
\begin{figure*}[h]
\centering
\includegraphics[width=\textwidth]{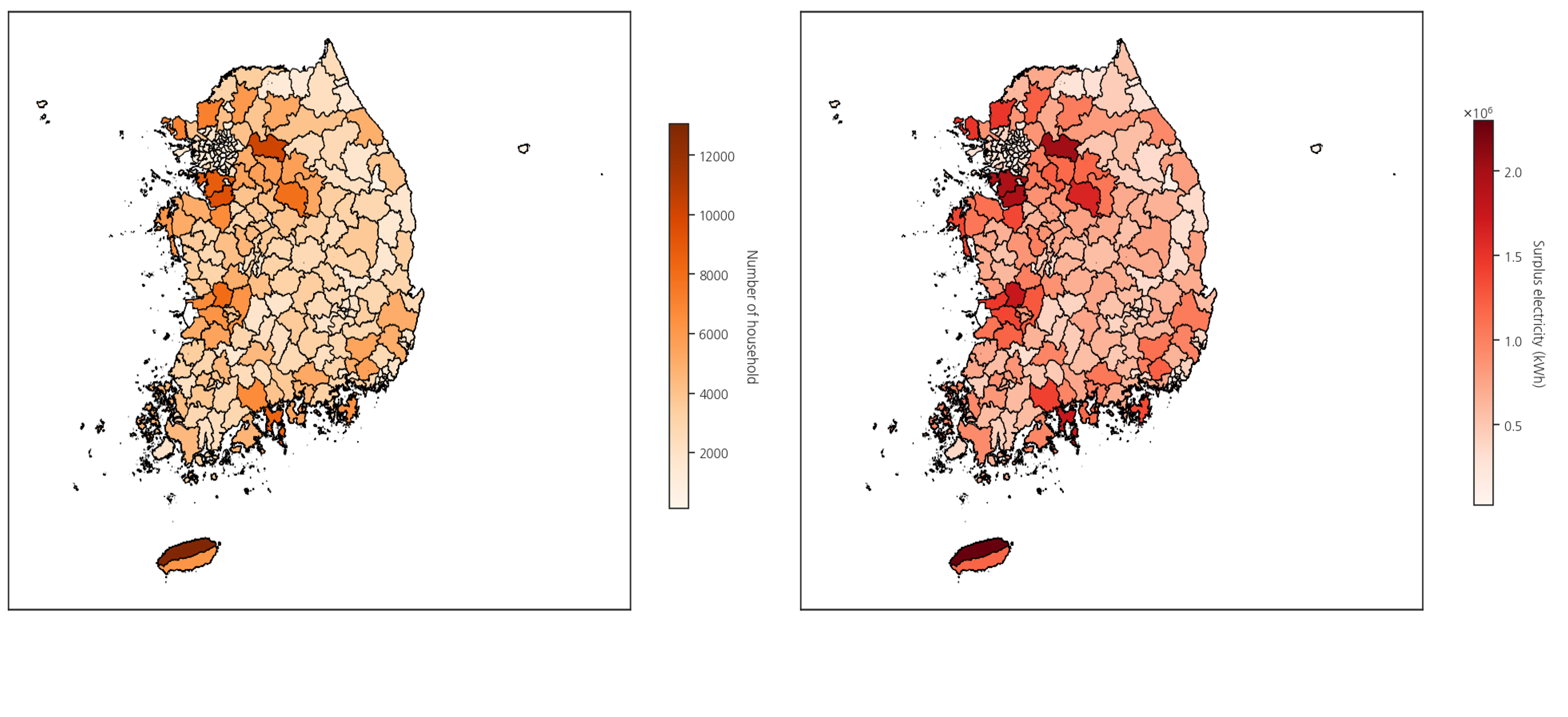}
\caption{Average surplus electricity in South Korea, 2024}
\label{fig:surplus_map}
\end{figure*}

The surplus electricity dataset was obtained from the KEPCO open data platform\footnote{\url{https://www.data.go.kr/data/15111708/fileData.do} (in Korean)}. It reports the amount of excess electricity generated by net-metering customers, aggregated by legal district (beopjeong-dong\footnote{``Beopjeong-dong'' refers to a legal neighborhood designation in South Korea used for administrative and statistical purposes. It does not always correspond to the actual administrative unit (``Haengjeong-dong'') that manages local affairs.}) and provided monthly and yearly. The dataset includes the number of customers and their corresponding surplus generation (kWh) for each administrative unit, as seen in Figure~\ref{fig:surplus_map}. Households with solar panel installations are prevalent in Gyeonggi Province near Seoul, along the southern coast, and on Jeju Island, suggesting a significant surplus of electricity. Jeju City in Jeju Special Self-Governing Province had the highest amount at 2,298,161.42 kWh, followed by Yangpyeong County in Gyeonggi Province at 2,026,834 kWh and Pyeongtaek City at 1,961,221 kWh. The lowest consumption was recorded in Dong-gu, Incheon Metropolitan City at 20,644 kWh, followed by Jung-gu, Seoul Special City at 29,641 kWh, and Jung-gu, Daegu Metropolitan City at 32,854 kWh. 
This surplus represents the residual power exported to the grid after self-consumption by distributed solar producers. Due to privacy regulations under Article 28 of the Personal Information Protection Act \citep{korea_pipa_article28}, data from regions with fewer than five customers were removed, resulting in missing values for sparsely populated or underdeveloped areas. This dataset captures the spatial distribution of small-scale renewable energy surpluses and forms the areal-level component of the multilevel optimization model.

\subsubsection{Official Land Price Data}
\begin{figure}[h]
\centering
\includegraphics[width=0.5\textwidth]{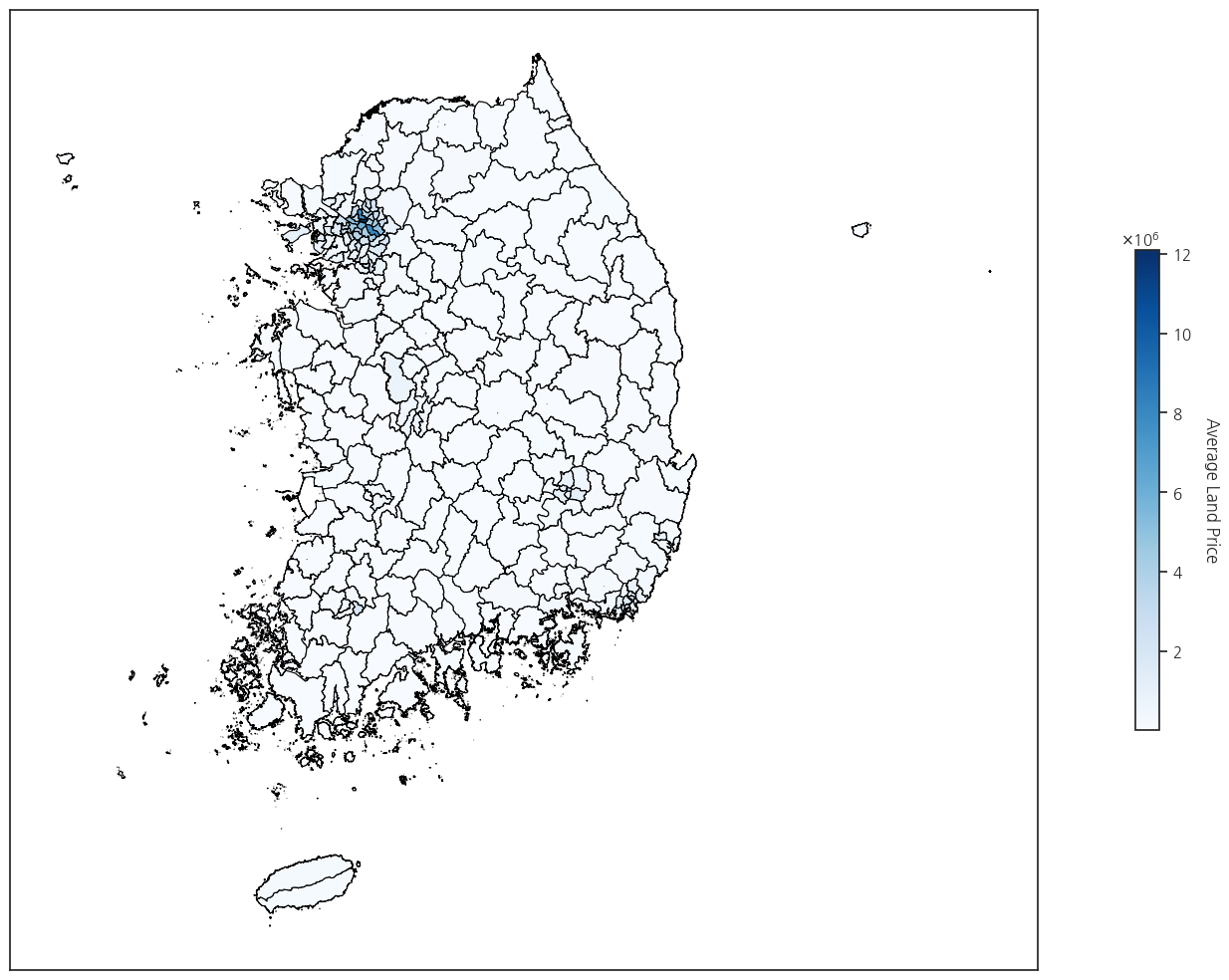}
\caption{Official land price}
\label{fig:officialvalue}
\end{figure}

The official land price dataset, published annually by the Ministry of Land, Infrastructure and Transport (MOLIT)\footnote{11, Doum 6-ro, Sejong City, South Korea, available at \url{https://www.molit.go.kr/english/intro.do}}, provides standardized appraised land values for individual parcels across South Korea. Each parcel’s price (KRW/m$^2$) is derived by comparing its physical and locational characteristics with those of officially designated reference lands. The valuation follows a regulated appraisal process: the standard land price, determined by the Minister of MOLIT, serves as a benchmark, and local governments adjust it using price adjustment coefficients based on land characteristics as defined in the official land valuation tables. The resulting land values are verified by licensed appraisers, reviewed by the local Real Estate Price Disclosure Committee, and finalized after collecting public opinions from landowners. Therefore, this dataset offers highly reliable, legally verified land value information that supports land taxation, real estate transactions, and policy formulation.
In Figure \ref{fig:officialvalue}, 
as expected, regions outside the capital city of Seoul showed little variation in land prices. The highest land value was recorded in Jung-gu, Seoul Special City at 11,760,951, followed by Gangnam-gu at 9,718,204 and Jongno-gu at 7,779,938. The lowest publicly announced land prices were recorded in Wando County, Jeollanam-do at 8,130, Shinan County at 8,135, and Yeongyang County, Gyeongsangbuk-do at 8,933.

\subsubsection{Topographic Slope Data}
The topographic slope data were derived from the Digital Elevation Model (DEM)\footnote{A Digital Elevation Model (DEM) is a 3D digital representation of a terrain's surface, providing elevation data in a grid format. The slope data in this study was derived by applying a first-order derivative to this DEM (i.e., calculating the rate of change in elevation).} provided by the Korea Institute of Geoscience and Mineral Resources\footnote{124, Gwahak-ro, Yuseong, Daejeon, South Korea, available at \url{https://www.kigam.re.kr/english/}}. The slope represents the rate of elevation change between adjacent cells on the Earth's surface, indicating terrain steepness. It is calculated as the maximum rate of change in elevation among neighboring grid cells, with higher values denoting steeper terrain and lower values indicating flatter surfaces. 

This dataset can be processed using commercial software such as ArcGIS or open-source GIS software such as QGIS. In this study, slope values were extracted from GeoTIFF files at a 30-meter spatial resolution, aggregated to grid cells matching the study’s analytical framework. The slope variable serves as a point-level constraint in the linear programming model, where steep areas receive penalty to ensure physical feasibility of mining site construction.
\begin{table}[h]
\centering \caption{Summary of Topographic Slope Data Structure} \label{tab:slope_detail_en}
\begin{tabular}{l l}
\toprule
\textbf{Attribute} & \textbf{Value} \\
\midrule
Data Format & GeoTIFF (GTiff) \\ Data Type & 32-bit floating point (float32) \\ Spatial Resolution & $30.0\text{m} \times 30.0\text{m}$ \\ Dimensions (H x W) & $20,027 \times 11,187$ pixels \\ Minimum Value & $0.00^\circ$ \\ Maximum Value & $77.89^\circ$ \\ Mean Value & $13.73^\circ$ \\
\bottomrule
\end{tabular}
\end{table}
Table \ref{tab:slope_detail_en} summarizes the detailed technical structure of this dataset. An average gradient of 13.73\% is a fairly steep hill. While gentler than typical residential stairs in everyday life, it is steep enough for shopping carts to roll down on their own \citep{steep}.

\begin{figure}[!ht]
\centering
\includegraphics[width=0.5\textwidth]{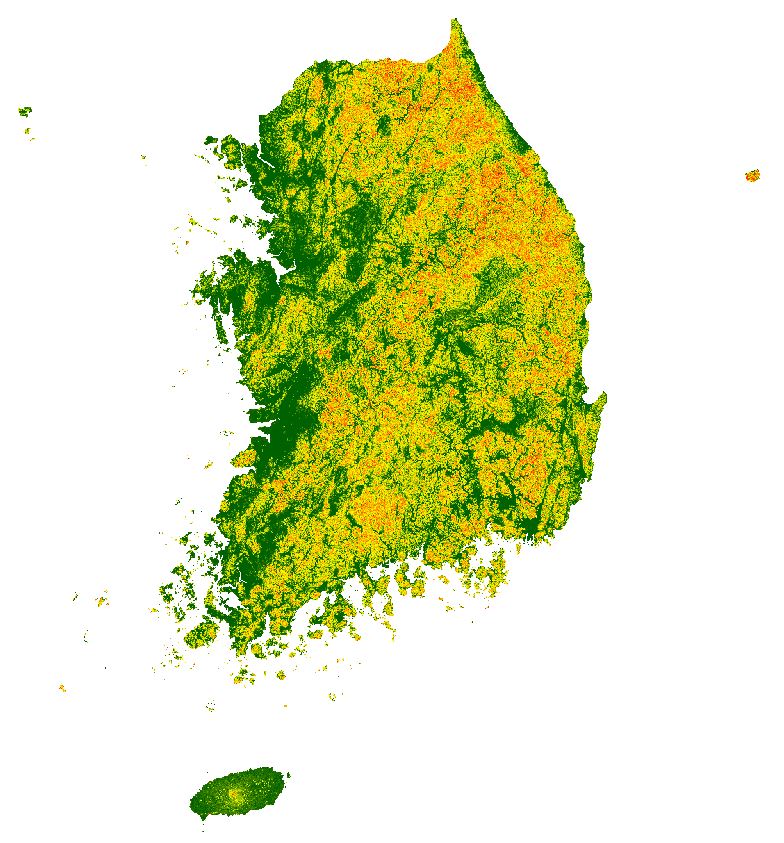}
\caption{Slope distribution map of South Korea derived from DEM data. Green regions indicate flat terrain, while red regions represent steeper slopes. This original image is from Korea Institute of Geoscience and Mineral Resources \citep{KIGAM2019SlopeAnalysis}.}
\label{fig:slope_map}
\end{figure}
As shown in Figure~\ref{fig:slope_map}, flatter regions are concentrated in western and southern parts of the country, while the eastern regions display more rugged terrain \citep{KIGAM2019SlopeAnalysis}. This spatial distribution was considered in the site suitability model since steep areas are less favorable for large-scale mining farm construction due to higher infrastructure and cooling costs.

\subsubsection{Bitcoin Network Data}
Bitcoin Network data were obtained from the Blockchain.com\footnote{York, United Kingdom, available at \url{https://www.blockchain.com/explorer}.}. The dataset includes the average network hash rate (TH/s), block reward BTC/block, and Bitcoin price (USD/BTC) for the study period. These variables are used to estimate expected revenue and to parameterize the profitability function within the linear programming formulation.

\subsection{Experimental Setup and Parameters}
\label{sec:parameters}
To ensure the reproducibility and objectivity of this analysis, the key fixed parameters used in the model were set as shown in Table \ref{tab:params_stage1}.

\subsubsection{Stage 1 (Areal) Cost-Benefit Analysis Parameters}
\label{stage1_setup}
The key parameters used in the detailed cost-benefit analysis for the Stage 1 regional optimization (Algorithm \ref{alg:stage1}) are as follows.

\begin{table}[h]
\footnotesize
\centering
\caption{Stage 1 Analysis Key Parameter Settings}
\label{tab:params_stage1}
\begin{tabularx}{\columnwidth}{@{} l l X l @{}}
\toprule
\textbf{Category} & \textbf{Parameter (Variable)} & \textbf{Value} & \textbf{Reference} \\
\midrule
Miner & Power per Machine& $4104.0$ $\text{kWh}$ & \citep{Bitmain2025_S21XP_Hyd} \\
\midrule
Area & Layout Mode & Container & \citep{HK3Spec} \\ 
& Miners per Container & $210$ units & \citep{HK3Spec} \\ 
& Area per Container & $51.7$ $\text{m}^2$ & \citep{HK3Install} \\ 
& Layout Margin Ratio & $30\%$ & \citep{zeus2025antspace} \\ 
\midrule
Land & Land Amortization & $20$ years & \citep{GHGProtocol2022}\\
\bottomrule
\end{tabularx}
\end{table}

The financial parameters for the infrastructure build-out are based on base-case assumptions derived from industry analyses of analogous projects, such as data centers and dedicated mining facilities due to the volatility of estimate, confidential information, and project size. Precise capital expenditure (CAPEX) quotations are proprietary and highly variable; therefore, we define a conservative baseline. We assume a Fixed CAPEX per Site of \$5 million representing the initial costs for site preparation, civil works, and the main substation \citep{bofa_whomakes_datacenter_2024_misc, thundersaidenergy_datacenters_economics_2024, mastt_datacenter_construction_2025}. The variable CAPEX per MW is assumed to be \$400,000 \citep{bofa_whomakes_datacenter_2024_misc, thundersaidenergy_datacenters_economics_2024}. This variable cost estimate is significantly lower than typical ``all-in'' data center benchmarks (which often range from \$7 million to \$15 million per MW) because it is explicitly defined to exclude the cost of the mining hardware (ASICs) itself. It is assumed to represent the procurement and installation costs of pre-fabricated, containerized solutions (such as the Bitmain AntSpace) and their direct utility hookups, rather than a full-scale ``brick-and-mortar'' data center build. Grid connection costs are based on the specific KEPCO tariff structure, which involves a step cost (Fee) of \$1.2 million for every $10.0\text{MW}$ step cost (Block) of new capacity \citep{bluecap2025}. For operational modeling, the infrastructure lifespan is set at 7 years, a conservative assumption for asset depreciation based on common tax regulations \citep{EINSNER2025}. The fixed OPEX per Site is assumed to be \$3 million per year, covering site-level costs such as staffing, security, and non-variable maintenance \citep{bluecap2025}. These baseline assumptions are later tested in the sensitivity analysis.

\subsubsection{Stage 2 (Point) Geospatial Screening Parameters}
\label{sec:point_params}
The parameters used for the Stage 2 screening, which searches for suitable sites within the regions selected in Stage 1, are as follows. For Topography, the maximum allowable slope ($\theta$) was set to $6.0^\circ$, based on United States Department of Housing and Urban Development guidelines \citep{HUDSlopeGuide}. For the search parameters, the unit site width ($W_j$) and unit site height ($H_j$) were both set to $80.0$ $\text{m}$ for `Yongin'. This value was derived by taking the square root of the required land area ($m^2$) for each site and rounding up to the nearest integer. Finally, a Search Stride ($\Delta$) of $20.0$ $\text{m}$ was used as part of the study's design to ensure search precision.

\subsection{Stage 1 (Areal) Optimal Region Selection Results}
\label{sec:stage1_results}
By applying the parameters from \ref{sec:parameters} to Algorithm \ref{alg:stage1}, the `Final adjusted net profit ($\Pi_{\text{adj}}$)' was derived for each scenario, increasing the number of regions $K$ from 1 to 5. This process aims to find the most efficient combination of national hubs.

\begin{figure}[h!]
\centering
\includegraphics[width=0.5\textwidth]{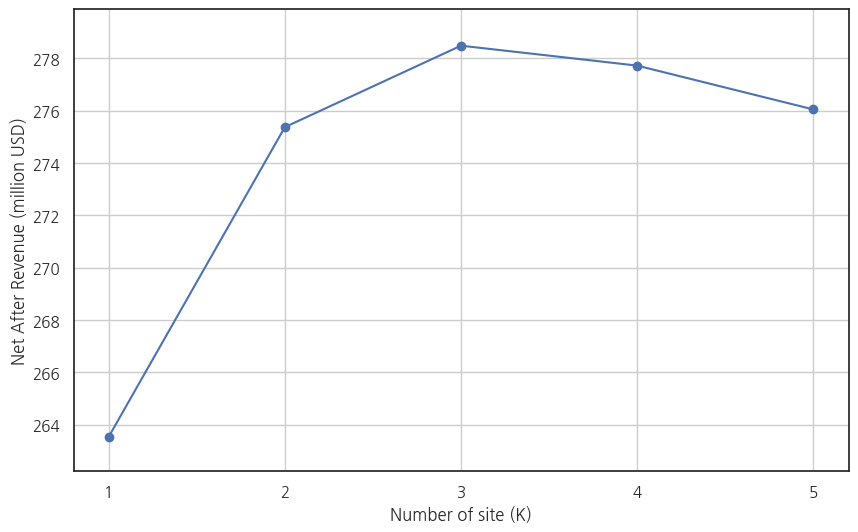}
\caption{Final adjusted net profit ($\Pi_{\text{adj}}$) by number of regions (K)}
\label{fig:k_vs_profit}
\end{figure}

\paragraph*{Determining the Optimal Number of Regions ($K^*$)} Figure \ref{fig:k_vs_profit} visualizes the change in final adjusted net profit as the value of $K$ increases. The analysis shows that for $K=1$ and $K=2$, net profit rose sharply, indicating high efficiency in securing surplus power relative to land and infrastructure costs. The maximum net profit ($\Pi_{\text{adj}}$) was reached at $K=3$ with \$ 278.5 million. For $K=4$ and $K=5$, the marginal benefit of securing new surplus power diminished, while the burden of fixed infrastructure costs (from $\mathcal{C}_{\text{opex}}$ and $\mathcal{C}_{\text{capex}}$) increased, leading to a decline in total net profit. Therefore, this study finalized $K^*=3$ as the optimal number of regions to maximize total net profit.

\begin{figure}[h!]
\centering
\includegraphics[width=0.45\textwidth]{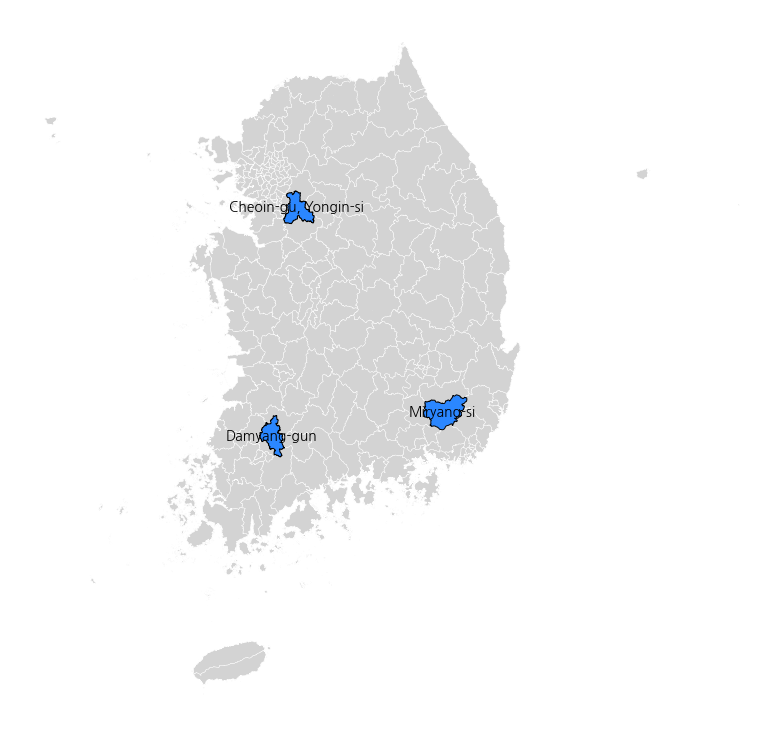}
\caption{Selected regions; Yongin, Damyang, and Miryang}
\label{fig:selected_regions}
\end{figure}

\paragraph*{Identification of Optimal Regions ($\mathbf{Y}^*$)}
Table \ref{tab:stage1_results_table} summarizes the optimal region combination ($\mathbf{Y}^*$) for $K^*=3$ and the detailed cost-benefit analysis for each region. The Stage 1 analysis identified `Yongin', `Miryang', and `Damyang' as the final selections, each representing a region with distinct economic and geographical characteristics, shown in  Figure~\ref{fig:selected_regions}.

The optimization process yielded a total of 40,471 mining units distributed across the three selected centers. Among them, \textbf{Yongin} accounts for the largest proportion with 18,978 units, followed by \textbf{Damyang} (11,853 units) and \textbf{Miryang} (9,640 units).

In terms of economic output, \textbf{Yongin} demonstrates the highest annual revenue of approximately \$164.8 million, resulting in a net annual profit of \$139.0 million after accounting for depreciation. \textbf{Damyang} and \textbf{Miryang} show comparable but slightly lower profitability levels, with annual net revenues of 91.0 million and \$76.9 million, respectively. The total annual net return across all centers is estimated at \$306.2 million.

From an energy perspective, the three centers collectively consume approximately 1.52 billion kWh per year, with \textbf{Yongin} again contributing the highest consumption (approximately 692 million kWh). These figures indicate a balanced trade-off between energy utilization and profitability, highlighting the model’s capability to identify spatially efficient yet economically viable mining configurations.

This implies that the Stage 1 effectively analyzed the complex trade-off between surplus power (revenue) and land and infrastructure (cost), deriving an optimal solution based on economic feasibility rather than merely ranking regions by available surplus power.

\begin{table*}[h]
\centering
\caption{Detailed Results of Optimal Regions ($K^*=3$) based on Net Profit} \label{tab:stage1_results_table}
{\begin{tabular}{l c c c c c}
\toprule 
\textbf{Region Code} & \textbf{Total Power} & \textbf{Fixed Machines} & \textbf{Annual Revenue} & \textbf{Annual Depreciation} & \textbf{Final Net Profit} \\ 
\textbf{} & \textbf{(GWh)} & \textbf{(Units)} & \textbf{($\Pi_{\text{Rev}}$, Million USD)} & \textbf{($D_K$, Million USD)} & \textbf{($\Pi_{\text{Net}}$, Million USD)} \\ 
\midrule 
Yongin & $692.09$ & $18,978$ & $164.78$ & $25.73$ & $\mathbf{139.05}$ \\ 
Damyang & $448.90$ & $11,853$ & $107.10$ & $16.07$ & $\mathbf{91.02}$ \\
Miryang & $376.10$ & $9,640$ & $89.99$ & $13.07$ & $\mathbf{76.92}$ \\ 
\midrule 
\textbf{Total ($K^*=3$)} & $\mathbf{1517.09}$ & $\mathbf{40,471}$ & $\mathbf{361.87}$ & $\mathbf{54.87}$ & $\mathbf{306.99}$ \\
\bottomrule
\end{tabular}}
\end{table*}

\subsection{Stage 2 (Point) Suitable Site Screening Results} \label{sec:stage2_results}
The three regions (Yongin, Damyang, Miryang) selected in \ref{sec:stage1_results} passed the macro-level economic screening based on their `Total Gross Area'. The objective of Stage 2 is to find the micro-level locations of `unit sites' where containers will actually be installed, which will then aggregate to fill this total area.

\paragraph*{Defining the Unit Site Area ($W_j \times H_j$)} The size of the Stage 2 sliding window, or `unit site area', was calculated based on the \texttt{fixed\_machines} for each region (from Stage 1) and the \texttt{mode=`container'} parameters (from \ref{stage1_setup}). This `Net Area' concept is distinct from the `Gross Area' ($2,500 \text{ m}^2/\text{MW}$) used in Stage 1.

For example, \textbf{Yongin} requires 18,978 machines. Applying the parameters from \ref{stage1_setup} (210 miners/cont., $51.7 \text{ m}^2$/cont., 30\% margin) yields: $$18,978 \text{ miners} / 210 \text{ (miners/cont)} \approx 91 \text{ containers}$$ $$91 \text{ (cont)} \times 51.7 \text{ (m}^2\text{/cont)} \times (1 + 0.30) \approx 6,116 \text{ m}^2$$ Based on this required net area ($6,116 \text{ m}^2$), we set the unit site window for the Stage 2 search to $80\text{m} \times 80\text{m}$ ($6,400 \text{ m}^2$). Table \ref{tab:stage2_results_summary} summarizes the calculated unit sizes and screening results for all three regions.

\begin{table}[h]
\centering
\caption{Summary of Stage 2 Suitable Site Screening Parameters and Results}
\label{tab:stage2_results_summary}
{\begin{tabular}{l c c c}
\toprule
\textbf{Region} & \textbf{Req. Net Area} & \textbf{Unit Site Size} \\
\midrule
Yongin & $6,116 \text{ m}^2$ (91 container) & $80\text{m} \times 80\text{m}$  \\
Damyang & $3,827 \text{ m}^2$ (57 container) & $65\text{m} \times 65\text{m}$ \\
Miryang & $3,090 \text{ m}^2$ (46 container) & $55\text{m} \times 55\text{m}$ \\
\bottomrule
\end{tabular}}
\end{table}

\paragraph*{Suitable Site Search Results} The sliding window search was executed on each region's slope raster ($slope_j$) using the unit site sizes from Table \ref{tab:stage2_results_summary} and the parameters from Sec \ref{sec:point_params} (max. slope $\theta=6.0^\circ$, stride $\Delta=20.0\text{m}$).
 
\begin{table}[h]
\centering
\caption{Number of Suitable Site Units}
\label{tab:suitable_site_search}
{\begin{tabular}{l c c c}
\toprule
\textbf{Region} & \textbf{Initial Unit Count} & \textbf{Available Unit Count} \\
\midrule
Yongin  & 147,533 & 19,456 \\
Damyang & 165,585 & 56,444 \\
Miryang & 210,333 & 67,522 \\
\bottomrule
\end{tabular}}
\end{table}

Table~\ref{tab:suitable_site_search} shows the initial unit count and available unit count for each region. The available count is the final available unit count derived from the initial unit count, reflecting the land use status data\footnote{A map detailing land use status by plot, categorised into six stages: urban, agricultural (paddy fields, dry fields), forestry (established, non-established), industrial, natural and cultural heritage, and reserved areas. The data is available at \url{https://www.vworld.kr/dtmk/dtmk_ntads_s002.do?svcCde=MK&dsId=30240} (in Korean), originally provided by National Geographic Information Institute.}. The results show that \textbf{Yongin} is the most constrained region for site selection, while \textbf{Damyang} and \textbf{Miryang} offer comparatively more, though still limited, availability.

Specifically, the land use data filtered out nearly 87\% of the initial candidate units in \textbf{Yongin} (reducing the count from 147,533 to 19,456). This indicates extensive existing development or conservation. \textbf{Damyang} and \textbf{Miryang} demonstrated greater availability, retaining 34.1~\% and 32.1~\% of their initial units, respectively. Despite starting with the largest number of potential sites (210,333), \textbf{Miryang} ultimately provides the highest count of available units (67,522) after accounting for present land use.\\

\begin{figure}[ht]
\centering
\includegraphics[width=0.5\textwidth]{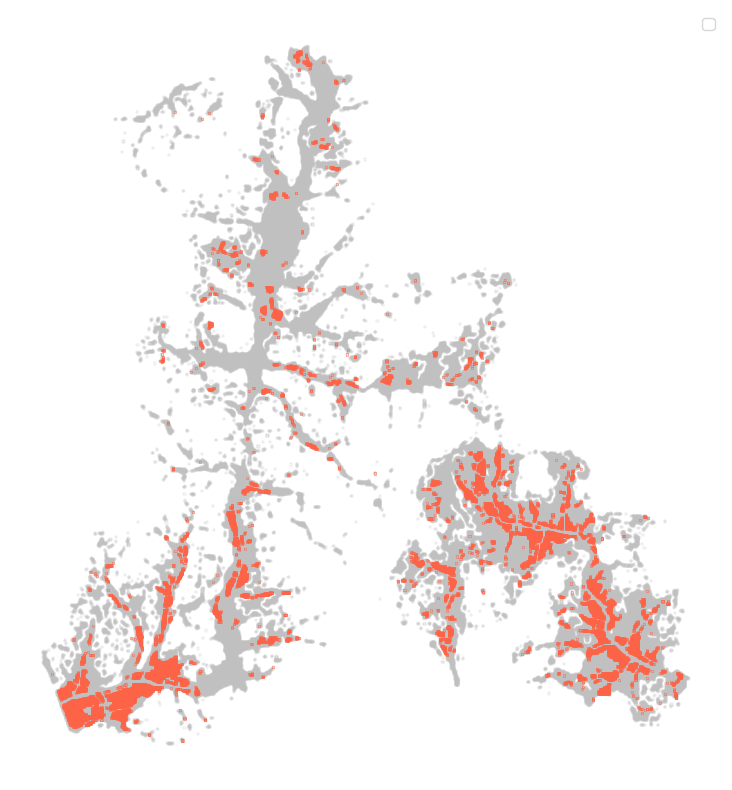}
\caption{Cheoin-gu, Yongin slope distribution and suitable sites} \label{fig:stage2_map_A}
\end{figure}

\paragraph*{Visualization of Suitable Site Distribution}
Figure~\ref{fig:stage2_map_A} visualizes the Stage 2 analysis for \textbf{Yongin (Cheoin-gu)}.
The vast number of initial candidate units (shown in grey, 147,533 units) identified by the slope constraint alone are spread across the district's flatter regions. However, after applying the land-use status filter, the final set of available units (shown in red, 19,456 units) is dramatically reduced by 87\%, the largest proportional constraint among the analyzed districts. This result underscores that while Yongin was selected for its high macro-level economic potential (Stage 1), its actual physical and regulatory avilability is severely constrained.
\begin{figure}[ht]
\centering
\includegraphics[width=0.5\textwidth]{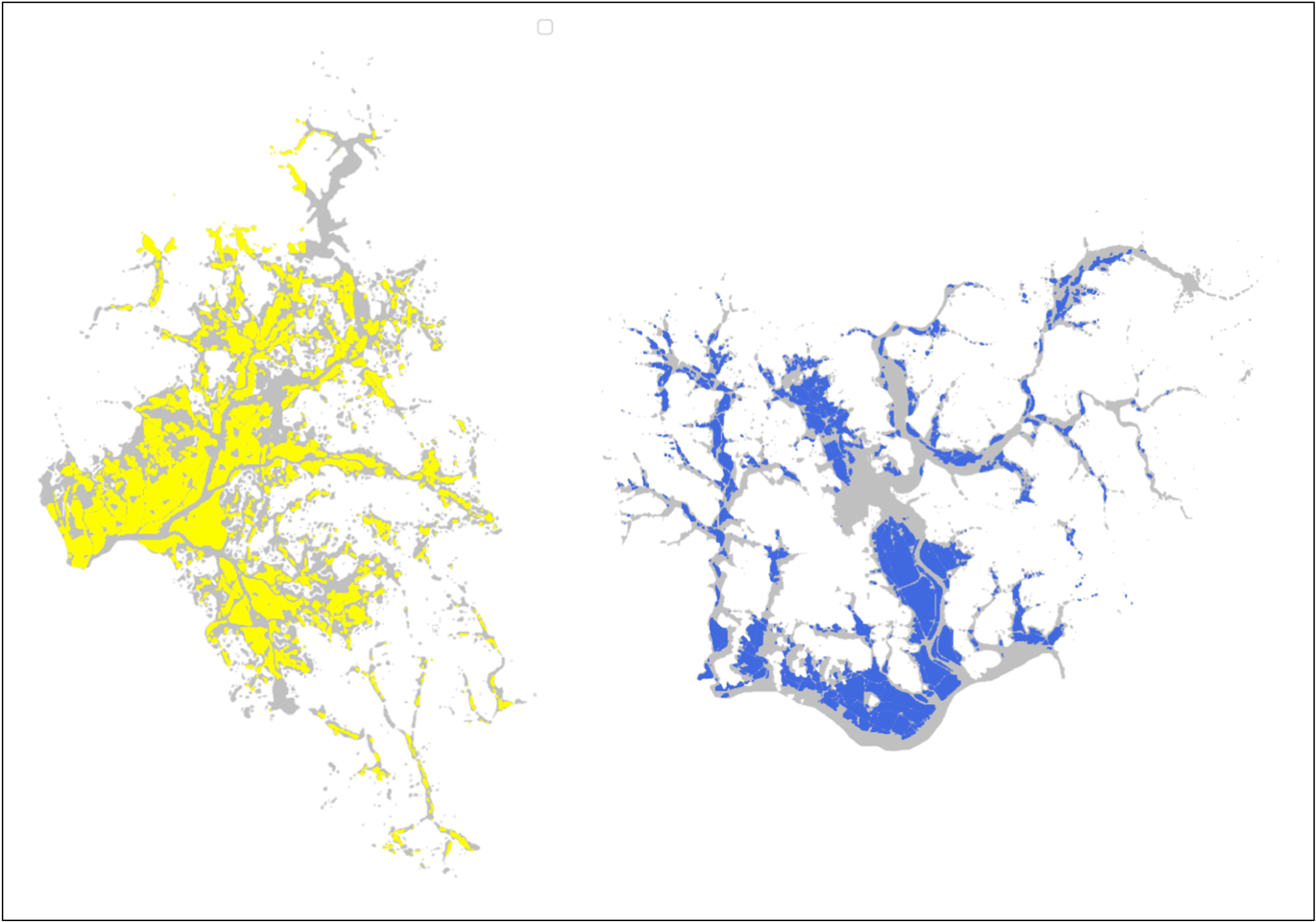}
\caption{Slope distribution and suitable sites (left: Damyang, right: Miryang)} \label{fig:combined}
\end{figure}
Figure \ref{fig:combined} shows the mining-eligible area units for Damyang and Miryang. The initial pool of candidate units in Damyang (shown in grey, 165,585 units) appears widely distributed across the district's river basins and flatter agricultural areas, as defined primarily by the slope constraint. However, imposing the land-use status filter significantly restricts this availability. The final set of deployable units (shown in yellow, 56,444 units) is dramatically reduced by approximately 66\%. This outcome reveals that, similar to Yongin, Damyang's high theoretical suitability is met with substantial on-the-ground constraints due to existing regulatory land classifications, significantly concentrating the final available areas.
Miryang initially presented the largest number of candidate units across the three regions, with 210,333 units (shown in grey) identified primarily by the slope constraint. These units were concentrated along the wide river valleys and low-lying plains. After the application of the restrictive land-use status filter, the final set of available units (shown in blue, 67,522 units) was constrained. Miryang experienced an approximate reduction of 68\%. This reduction emphasizes the trade-off between macro-level geographic suitability and the micro-level regulatory feasibility of siting new energy infrastructure within Miryang's administrative boundaries.

\begin{figure}[ht]
\centering
\includegraphics[width=0.5\textwidth]{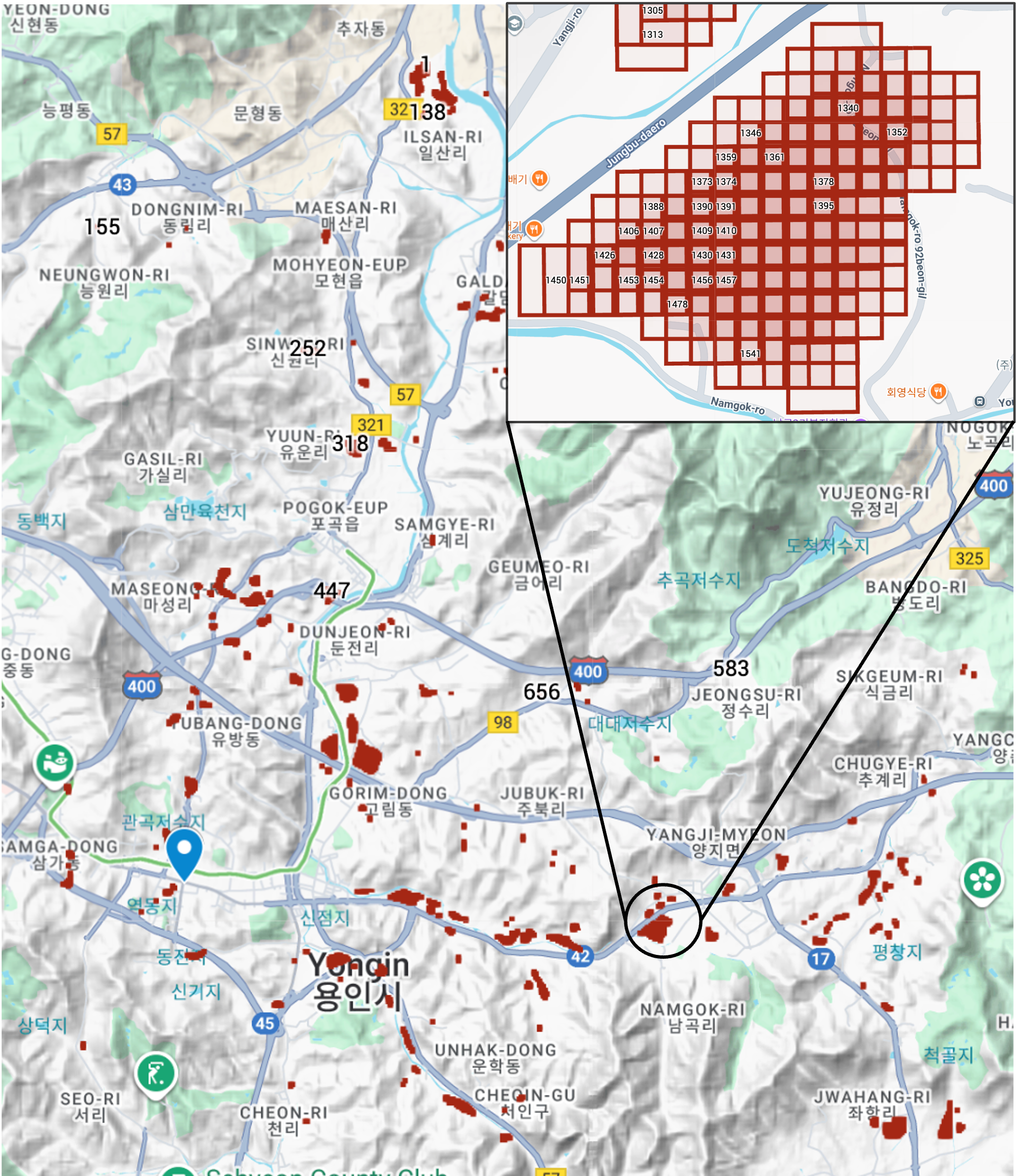}
\caption{Cheoin-gu, Yongin Slope Distribution and Suitable Sites in Google Map}
\label{fig:yongin_googlemap}
\end{figure}

To provide real-world context for the 19,456 available sites identified in Yongin, a sample of 2,000 units was visualized on the Google My Maps\footnote{Google My Maps is a web-based, cloud-hosted mapping service provided by Google. It allows users to create, customize, and share interactive maps. The platform facilitates the visualization of spatial data by enabling the import of various data formats (such as KML, CSV, or Google Sheets), the addition of custom placemarks, and the drawing of vector features (points, lines, and polygons) organized into distinct layers. Available at \url{https://www.google.com/maps/about/mymaps/.} } platform described in Figure \ref{fig:yongin_googlemap}.
This tool was chosen for its accessibility and its capability to render custom datasets as interactive map layers, allowing us to easily explore the study's geographic results. As seen in the background slope map, most of the region is unsuitable for construction despite its Stage 1 profitability. In the upper right corner of the figure, there is a zoomed-in map in the part of Yongin area. The suitable sites (red rectangles) that passed the $\theta=6.0^\circ$ constraint are highly clustered in specific fat areas near Yangji-myeon. The inset in the top-right corner magnifies one such cluster, demonstrating the physical $80\text{m} \times 80\text{m}$ grid scale of the individual unit sites that passed both the slope and land-use screening by accessing the live map at the following URL: \url{https://www.google.com/maps/d/u/4/edit?mid=19UOXt_t_GcNSR2TnslBt3X0G-WvDUqg&usp=sharing}.

\section{Conclusion}
\label{sec:ch4}
This thesis confronted the dual challenge of managing distributed renewable energy surplus in South Korea and the spatially dependent, energy intensive nature of Bitcoin mining. We identified a critical gap in existing site-selection literature. To bridge this methodological gap, we developed and implemented a Two-Stage Hierarchical Optimization framework. This framework integrates macro-level economic optimization with micro-level physical site screening. 

Furthermore, the significance of this research extends beyond mere economic optimization and grid management. The high energy consumption of Bitcoin, often criticized as a liability, is being re-contextualized as a strategic imperative. As theorists like U.S. Space Force Major Jason Lowery argue in `Softwar', Proof-of-Work is not simply an industry but a novel form of ``power projection'' that converts raw electrical energy into immutable digital security and strategic deterrence \citep{Blockmedia2023Softwar, Lowery2023Softwar}. In this ``Softwar'' framework, a nation's ability to secure the network through energy expenditure translates directly to influence and sovereignty in the digital age.

From this strategic perspective, South Korea's abundant but `curtailed' net-metering surplus power is not a waste product, but a critical, untapped strategic resource. This thesis provides the first actionable, data-driven roadmap to translate that distributed energy potential into a robust, national-scale infrastructure. Our framework moves beyond theory, offering a practical methodology for South Korea to operationalize this capability, effectively transforming a grid-balancing challenge into an opportunity to secure digital sovereignty.

In Stage 1 (Areal-Level Optimization), we moved beyond simple suitability ranking by implementing a detailed cost-benefit analysis (Algorithm \ref{alg:stage1}). This model optimized for a Final Adjusted Net Profit by balancing revenue from surplus solar power against comprehensive, non-linear costs, including annualized land acquisition and stepped infrastructure CAPEX. Analysis across $K=1$ to $K=5$ candidate sets, revealed that $K=3$ provides the most robust and balanced solution for maximizing annual profitability, specifically in Yongin, Damyang, and Miryang, yielded the maximum potential net profit of approximately \$307.0 million (Table~\ref{tab:stage1_results_table}). This result demonstrated a clear trade-off between securing surplus power and the diminishing returns from fixed infrastructure costs at $K=4$ and $K=5$ (Figure~\ref{fig:k_vs_profit}). 

Subsequently, Stage 2 (Point-Level Screening) translated this macro-level economic potential into micro-level physical feasibility. By applying a sliding window search (Section~\ref{sec:stage2}) with topographic constraints (max slope $\theta=6.0^\circ$) and land-use filtering, we identified the specific, constructible `unit sites' within the optimal regions. The results underscored the critical importance of this hierarchical step: while \textbf{Yongin} was the most profitable region in Stage 1, the Stage 2 analysis revealed it to be the most topographically and spatially constrained, with nearly 87\% of its initial candidate sites failing the land-use screen (Table~\ref{tab:suitable_site_search}). This process successfully generated a final list of viable, geolocated candidate sites (Figure~\ref{fig:yongin_googlemap}), bridging the gap from regional potential to actionable parcels.

This research makes two primary contributions. Methodologically, the Two-Stage Hierarchical Optimization framework provides a robust and scalable alternative to traditional GIS-MCDA approaches. By separating areal-level economic optimization from point-level physical screening, it effectively integrates heterogeneous spatial data and overcomes the limitations of subjective weighting. Practically, this study offers a concrete, data-driven strategy for South Korean energy policy. It provides KEPCO and private operators with a validated model to co-locate flexible loads (like Bitcoin mining) to monetize otherwise curtailed net-metering surplus, thereby improving grid efficiency and supporting renewable energy integration.

Several limitations must be acknowledged. The analysis relied on conservative, estimated costs for CAPEX and OPEX, which are highly variable in practice (Section~\ref{sec:parameters}). The surplus electricity data, while official, contained omissions due to privacy regulations (Section~\ref{sec:surpluselectricity}). Furthermore, our model assumed a zero-cost energy input based on surplus utilization and did not explicitly model the variable cost of grid interconnection (e.g., distance to substations), noise regulations, or water availability for cooling, which are critical for final project development.

Future research should focus on refining this framework by incorporating more granular data. This includes integrating detailed grid infrastructure maps to model interconnection costs as a variable, rather than a fixed, cost. A dynamic temporal analysis, matching hourly surplus data with real-time mining profitability, would also enhance the model's precision. Finally, the model's adaptability could be tested by applying it to other energy-intensive flexible loads (e.g., data centers, hydrogen electrolysis) or to different geographic regions facing similar renewable curtailment challenges.

\ifCLASSOPTIONcaptionsoff
  \newpage
\fi

\bibliographystyle{abbrvnat}
\bibliography{ref}

\end{document}